\documentclass[12pt,preprint]{aastex}

\shorttitle{A Moving Substructure in A Cluster of Galaxies}
\shortauthors{Takizawa}

\begin{document}


\title{Hydrodynamic Simulations of A Moving Substructure in A Cluster of
       Galaxies: Cold Fronts and Turbulence Generation}


\author{M. Takizawa}
\affil{Department of Physics, Yamagata University, 
       Yamagata 990-8560, Japan}
\email{takizawa@sci.kj.yamagata-u.ac.jp}



\begin{abstract}
 We perform three dimensional hydrodynamical simulations of a moving
 substructure in a cluster of galaxies. We investigate dynamical
 evolution of the intracluster medium (ICM) in and
 around the substructure moving radially in the larger cluster's
 gravitational potential, and its observational consequences. After the
 substructure
 passes the larger cluster's center, a bow shock and clear contact
 discontinuity form in front of it. The contact discontinuity looks like
 a sharp cold front in the X-ray image synthesized from the
 simulation results. This agrees with 
 a structure found in 1E 0657-56. The flow structure
 remains laminar before the first turnaround because the ram-pressure
 stripping is dominant over the development of Kelvin-Helmholtz instabilities
 on the boundary between the substructure and the ambient ICM.
 When a subcluster oscillates radially around the larger cluster's
 center, both Kelvin-Helmholtz and Rayleigh-Taylor instabilities develop
 well and the flow structure becomes highly turbulent. Around the turnaround,
 the subcluster's cold gas is pushed out of its potential well. 
 Therefore, the cold gas component appears to be in front of the
 subcluster. A relatively blunt cold front appears in the simulated
 X-ray image because the contact discontinuity is perturbed by Rayleigh-Taylor
 instabilities. This can explain the ICM structure found in A168.
\end{abstract}



\keywords{galaxies: clusters: general --- galaxies: clusters: individual 
          (\objectname{A168}, \object{1E 0657-56}) --- hydrodynamics ---
          intergalactic medium --- X-rays: galaxies: clusters }


\section{Introduction}

Clusters of galaxies are the largest virialized objects in the present
universe. In the standard scenario of cosmological structure formation,
larger structures form more recently. Thus, clusters of galaxies 
are recognized as
the virialized objects that form in the most recent epoch, and they
are structure formation sites that we can observe in detail at
relatively low redshifts. 
In fact, some clusters are forming now, which is evident from moving
substructures found through X-ray observations
\citep[e.g.,][]{Mark00,Mark02,Vikh01b}.

Moving substructures cause various characteristic structures in the
intracluster medium (ICM). A bow shock and contact discontinuity will
form in front of them. The latter most likely corresponds to a ``cold
front'' that is a sharp feature in the X-ray brightness distribution 
found by {\it Chandra} in a number of clusters
\citep[e.g.,][]{Mark00,Vikh01b,Sun02,Kemp02,Dupk03}. 
Moving substructures most likely generate turbulence in the ICM through fluid 
instabilities. \cite{Mazz02} found arc-like filamentary
features in the X-ray image of A3667, 
which are probably produced via hydrodynamic instability. 
Turbulence plays crucial roles in cluster evolution. It may
have a significant impact on the transport and mixture of the heavy
elements and 
thermal energy. Fluid turbulence generates magnetic turbulence,
which accelerates non-thermal particles and causes various high energy
phenomena in the intracluster space \citep[e.g.,][]{Ohno02,Fuji03,Brun04}.
{\it Astro-E2} satellite, which is
planed to be launched in 2005, will detect broadened lines due to the
turbulent motion \citep[][]{Suny03,Fuji05,Pawl05}. 
In order to study above-mentioned
issues, it is crucial to clarify the generation processes and structure
of the ICM turbulence.

Numerical simulations are necessary to investigate the ICM structure and
evolution in detail. In fact, highly turbulent ICM
and structures similar to cold fronts have been found in N-body +
hydrodynamical simulations with cosmological initial conditions 
\citep[e.g.,][]{Bial02,Naga03,Suny03,Math05,Titt05}.
It is true that such
simulations are useful in order to investigate ICM evolution as a whole
in realistic situations. In such simulations, however, it is often difficult 
to derive clear physical interpretations, to control initial conditions
for some purposes, and to achieve very high resolution in particular
interesting regions efficiently.
Therefore, simulations with a little bit less realistic but idealized
and well-controlled conditions are complementary and useful. 
For example, some study cluster mergers 
\citep[][]{Roet96,Taki99,Taki00,Rick01,Ritc02}, 
and others deal with ICM dynamics near moving
substructures \citep[][]{Hein03,Asai04}.
Such simulations are suitable in order to clarify physical
processes concerned, compare simulation results with specific
observations, achieve very high resolution efficiently 
in the region concerned, and so on.

In this paper, we adopt the latter approach in order to follow flow
evolution in and around a moving substructure with high spatial
resolution. This is necessary especially to investigate the structure and
generation processes of the turbulence. Because the field of view of 
{\it Astro-E2} XRS is not so large, detailed flow velocity structure
information obtained from numerical simulations is very important 
to predict (and/or interpret) {\it Astro-E2} XRS
observations and study ICM dynamics. In particular, information 
about when and where turbulence occurs is useful.
We perform three dimensional hydrodynamical simulations of a moving
substructure in a cluster of galaxies. 
To emulate the time-dependent environment near
the moving subcluster in the larger cluster, we take account of 
the time evolution of the boundary conditions in front of the subcluster.
We calculate the subcluster's position and velocity relative to the
larger cluster  
with an approximation that the subcluster is a test particle
in the larger cluster's gravitational potential. We take this
information into account calculating the boundary conditions in front of the
subcluster in our hydrodynamical simulations.

The rest of this paper is organized as follows. In \S \ref{s:simu} 
we describe the adopted numerical methods and initial conditions 
for our simulations. In \S \ref{s:resu} we present the results. 
In \S \ref{s:turb} we discuss a turbulence generation scenario by substructure
motion based on our simulation results. In \S \ref{s:comp} we compare
our results with X-ray observations of 1E 0657-56 and A168.
In \S \ref{s:summ} we summarize the results and discuss their implications.

\section{The Simulations}
\label{s:simu}

\subsection{Numerical Methods}
In the present study, we use the Roe TVD scheme to follow the dynamical
evolution of the ICM \citep[see][]{Hirs90}. 
The Roe scheme is a Godunov-type method with a linearized
Riemann solver \citep[][]{Roe81}.
It is relatively simple and good at capturing shocks without any
artificial viscosity. Using the MUSCLE approach
and a minmod TVD limiter, we obtain second-order accuracy without any
numerical oscillations around discontinuities. To avoid negative
pressure, we solve the equations for the total energy and entropy
conservation simultaneously. This method is often used in astrophysical
hydrodynamical simulations where high Mach number flow
can occur \citep[][]{Ryu93,Wada01}.

We focus on the flow evolution in the region near the subcluster moving
in the larger cluster. We calculate the evolution of the subcluster's 
position and velocity relative to the larger cluster with an
approximation that subcluster is a test particle in the larger cluster's
gravitational potential. Only gravitational interaction is considered
between them. 
The larger cluster's potential is fixed in space. The subcluster's
potential is fixed in shape, with the center of the 
subcluster's potential moving in the potential of the larger cluster as
a test particle would.  
The gas is not self-gravitating so simply responds to the sum 
of the larger cluster and subcluster potentials.
We set a simulation box in a frame comoving with the
subcluster. The coordinate system is set so that the subcluster's center
is always at the origin and the larger cluster's ICM approaches the subcluster
from the $+x$ side initially.
Hydrodynamic calculations are performed only inside this box.
Taking account of the position and velocity relative to the larger
cluster, we make the boundary conditions in front of the subcluster
change with time. Free boundary conditions are adopted for the other
boundaries. We assume that the presence of the subcluster does not affect 
the properties of the larger cluster outside the simulation box 
such as its gravitational potential shape, ICM density structure, and
ICM temperature structure. Gradients of the larger cluster's 
physical quantities in the direction perpendicular to the subcluster's
motion are neglected inside the simulation box.

\subsection{Models and Initial Conditions}
We assume that both the subcluster and larger cluster are represented by
a conventional isothermal beta-model in the initial state. 
The gravitational potential distribution is that of a King model for
each cluster. The density distribution and mass inside the radius $r$ 
of a King model are
\begin{eqnarray}
   \rho(r) &=& \rho_0 (1+x^2)^{-3/2},  \\
   M(r) &=& 4 \pi \rho_0 r_{\rm c}^3 [ 
         \ln \{ x + (1+x^2)^{1/2} \} - x (1+x^2)^{-1/2}
              ],
\end{eqnarray}
where $x=r/r_{\rm c}$. 
Here, $\rho_0$ and $r_{\rm c}$ are the central density and core radius,
respectively. The gravitational mass distribution is assumed to be dominated 
by the dark matter (DM) component.
We set the outer boundary of the DM distribution to $5 r_{\rm c}$. Thus,
the total mass $M$ can be used as a model parameter, instead of the
central density $\rho_0$. We set $r_{\rm c} = 200$ kpc and 
$M=8.57 \times 10^{14} M_{\odot}$  for the larger cluster, and 
$r_{\rm c} = 40$ kpc and $M=1.43 \times 10^{14} M_{\odot}$  for the subcluster.
The ICM is assumed to be in hydrostatic equilibrium with isothermal
temperature distribution within the DM potential. In the isothermal 
beta-model, the gas density distribution is
\begin{eqnarray}
  \rho_{\rm g}(r) = \rho_{\rm g,0} (1+x^2)^{-3 \beta / 2},
\end{eqnarray}
where $\rho_{\rm g,0}$ and $\beta$ are the central gas density and
the specific energy ratio of the dark matter to the gas, respectively.
We set the initial temperature so that $\beta = 0.8$ for each cluster.
The mass density of the gas is assumed to be a tenth of that of the DM
in the center. The size of the simulation box and the number of the grid
points are (800 kpc)$^3$ and $400^3$, respectively.

\begin{figure}
\epsscale{.70}
\plotone{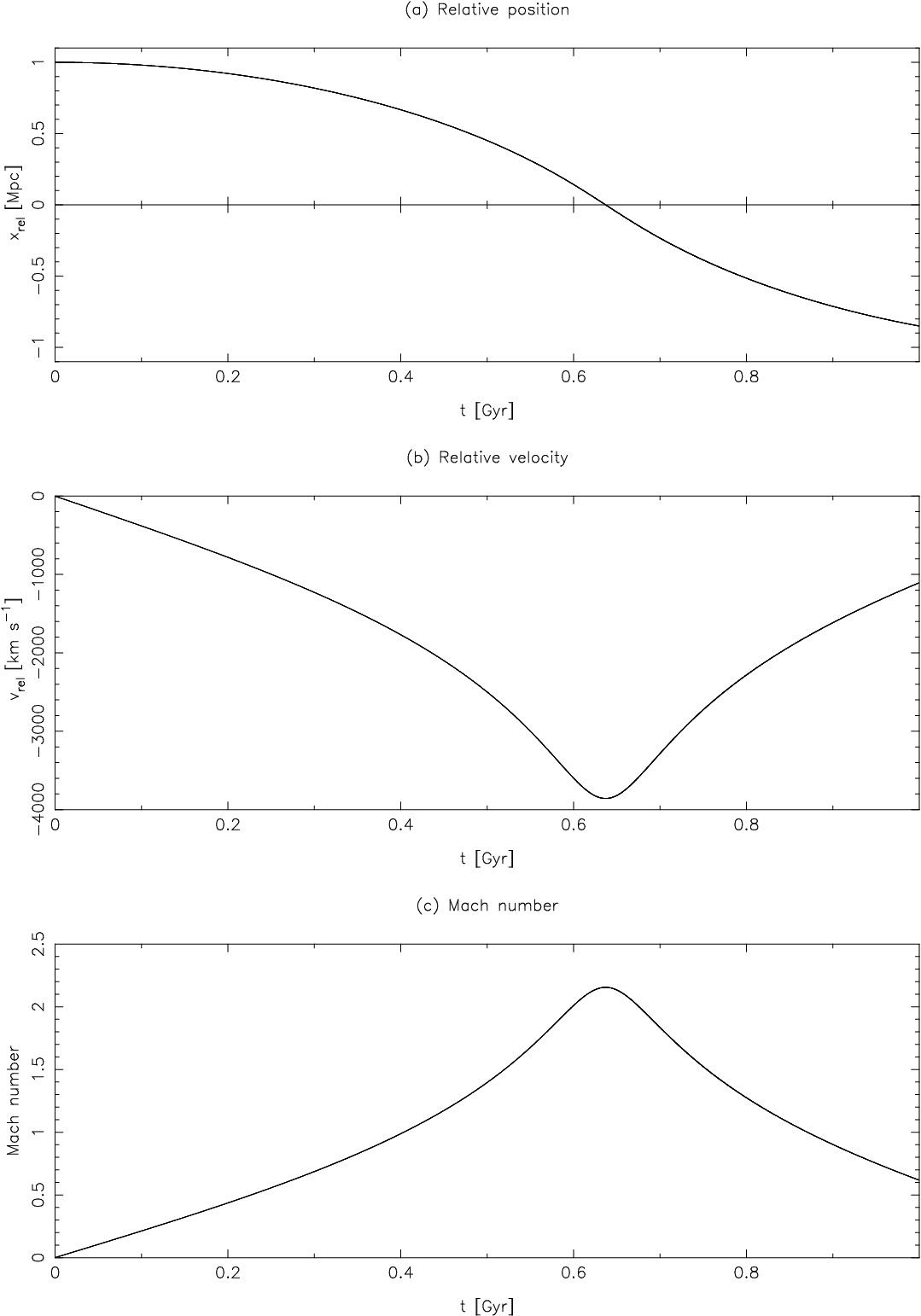}
\caption{Time evolution of the subcluster's (a) position, (b) velocity,
 and (c) Mach number relative to the larger cluster's center for the radial
 infall model.}
\label{fig:relxrelv1}
\end{figure}

\begin{figure}
\epsscale{.70}
\plotone{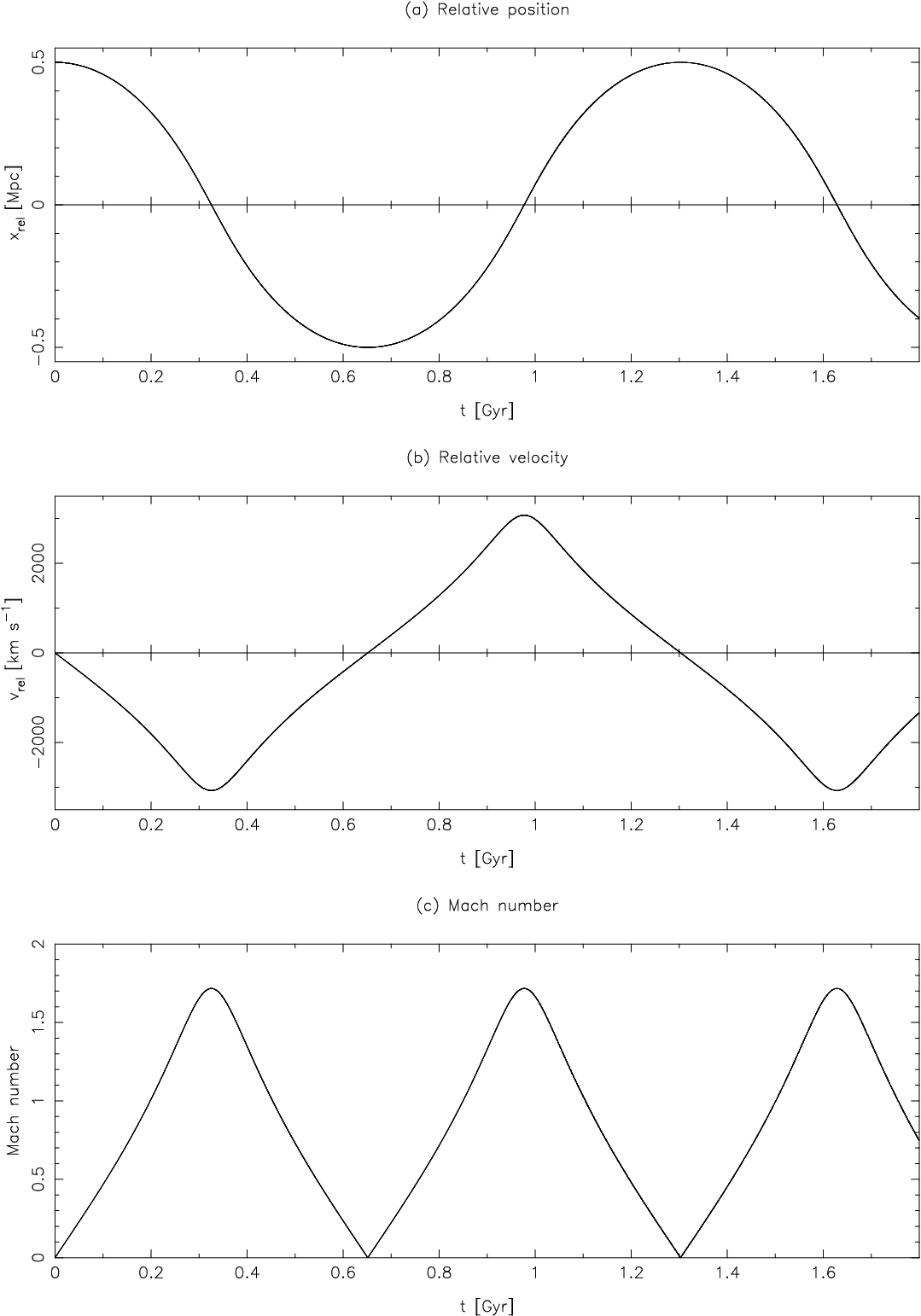}
\caption{The same as figure \ref{fig:relxrelv1}, but for the sloshing model.}
\label{fig:relxrelv2}
\end{figure}

We show the results for two representative cases. In one case, 
the subcluster falls radially into the larger cluster from the outskirts,
crosses the center, and reaches the opposite side of the larger cluster.
We call this ``the radial infall model''. In the other case, the subcluster 
is oscillating near the bottom of the larger cluster's gravitational
potential well. We call this ``the sloshing model''.
Figures \ref{fig:relxrelv1} shows the time evolution of the subcluster's
position, velocity, and Mach number relative to the larger cluster 
for the radial
infall model. In this model, the subcluster is at rest 1 Mpc away from
the larger cluster's center in the initial state. 
We follow the subcluster's motion for 1
Gyr until it reaches about 0.8 Mpc away from the center on 
the opposite side of the initial position. 
Figures \ref{fig:relxrelv2} shows the same as figure
\ref{fig:relxrelv1}, but for the sloshing
model. In this model, the subcluster is at rest 0.5 Mpc
away from the center initially. We follow its motion for 1.8 Gyr.
The subcluster crosses the larger cluster's center three times.

\section{Results}
\label{s:resu}

\begin{figure}
\includegraphics[angle=270,scale=0.7]{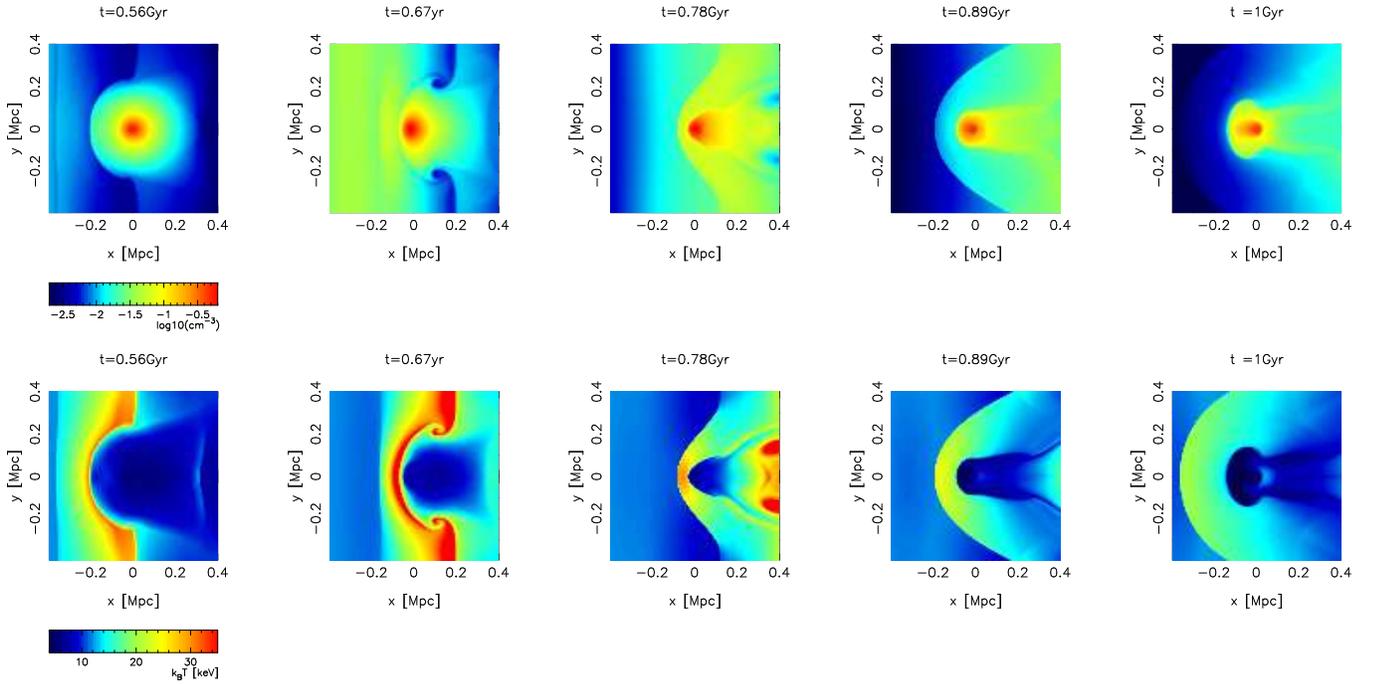}
\caption{Upper panels show snapshots of the density distribution on the
 $z=0$ surface at $t=0.56$, $0.67$, $0.78$, $0.89$, and $1.0$ Gyr of the
 radial infall model. Lower panels show the same ones but for the 
 temperature distribution.}
\label{fig:dentemp1}
\end{figure}

Figure \ref{fig:dentemp1} shows snapshots of the density and temperature
distribution on the $z=0$ surface at $t=0.56$, $0.67$, $0.78$, $0.89$,
and $1.0$ Gyr of the radial infall model. The upper and lower panels show
the density and temperature distribution, respectively. The subcluster
starts to fall at $t=0$, and is accelerated towards the larger cluster center. 
Both the relative velocity and the density in front of the subcluster are
increasing. Therefore, the subcluster suffers increasing ram
pressure, which is the product of the density and the square of the
relative velocity. As a result, the subcluster's ICM is gradually stripped
off in the outer region. The forward and reverse shocks are seen in the
density and temperature distribution at $t=0.58$ and $0.67$ Gyr. The
reverse shock cannot penetrate into the substructure's core. Thus,
a cold remnant survives. The temperature becomes higher 
between the two shocks, especially just behind the reverse
shock. The subcluster reaches the larger cluster's
center at $t \sim 0.64$ Gyr. Small eddy-like structures develop on the
boundary between the substructure and the surrounding gas through
Kelvin-Helmholtz instabilities.
After the substructure passes the larger cluster's center, the cold remnant
of the substructure is decelerating and enters the lower density region.
Therefore, it suffers the decreasing ram pressure and no more gas is
stripped off. A bow shock and contact discontinuity are clearly seen
in front of the remnant.
Although the small eddies are generated through Kelvin-Helmholtz
instabilities, the flow structure is basically laminar in the radial
infall model.

\begin{figure}
\includegraphics[angle=270,scale=0.7]{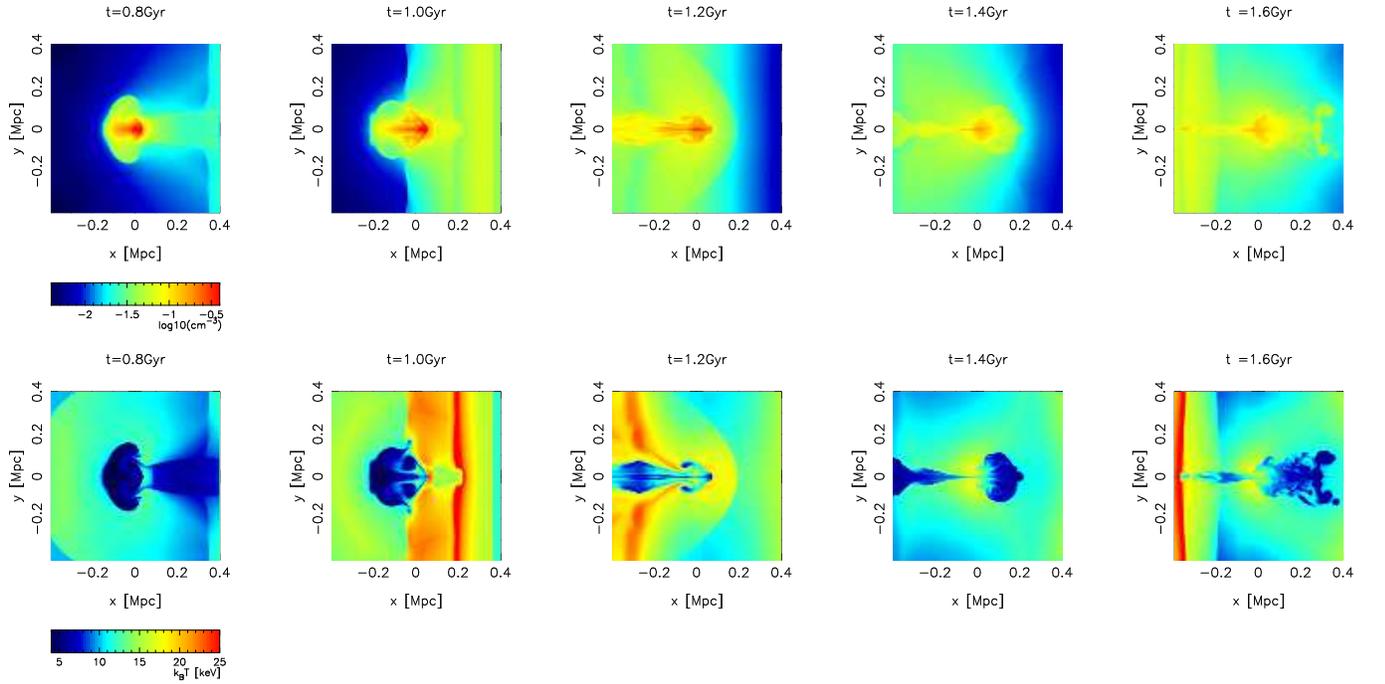}
\caption{The same as figure \ref{fig:dentemp1}, but for the sloshing
 model. The snapshots at $t=0.8$, $1.0$, $1.2$, $1.4$, and $1.6$ Gyr
 are presented.}
\label{fig:dentemp2}
\end{figure}

Figure \ref{fig:dentemp2} shows the same as figure \ref{fig:dentemp1},
but for the sloshing model. The snapshots at $t=0.8$, $1.0$, $1.2$, $1.4$, 
and $1.6$ Gyr are presented. In this model, the subcluster is at rest 
0.5 Mpc away from the larger cluster's center in the initial state.
It first passes the center at $t \sim 0.33$ Gyr, reaches 0.5 Mpc away from
the center on the opposite side, and turns around at $t \sim 0.65$ Gyr.
Then it falls back again, cross the center at $t \sim 0.98$ Gyr, reaches
the initial position at $t \sim 1.3$ Gyr, and turns around again. 
Figure \ref{fig:dentemp2} shows the evolution from the first
turnaround to the second one. At $t=0.8$ Gyr the subcluster has already
passed the larger cluster's center and turned around. A mushroom-like
structure forms because of the ram pressure and Kelvin-Helmholtz
instabilities. Furthermore, another mushroom-like structure forms
inside the first one after the turn around. After the second passage of
the main cluster's center, a bow shock forms again in front of 
the substructure ($t=1.2$ Gyr). 
Acceleration toward the the larger cluster center acts on the substructure 
in the larger cluster's potential. This means
that the inertial force in the $+x$ direction 
as well as the gravitational force of the subcluster act on the gas 
inside the subcluster in the frame comoving with the subcluster. 
If the former is stronger than the latter, 
the effective gravity is in the $+x$ direction.
Therefore, the gas in the region of $x>0$ is accelerated towards the
outside of the subcluster potential, rather than its center.
Because the denser gas is located nearer the center, 
Rayleigh-Taylor instabilities develop at the bow shock and contact
discontinuity. As a result,
cold gas seems to be in front of the subcluster's center ($t=1.4$ Gyr).
Finally, the cold cloud breaks into small pieces and the flow becomes
highly turbulent ($t=1.6$ Gyr).

\section{A Turbulence Generation Scenario}
\label{s:turb}

Let us discuss a turbulence generation scenario by moving substructures
in clusters of galaxies based on the results mentioned above. 
When a subcluster falls into a larger cluster
almost radially, a situation similar to the radial infall
model is realized before the first turnaround. The gas in the outer 
region is stripped off because of the ram pressure, but Kelvin-Helmholtz
instabilities hardly develop on the boundary. Therefore, the flow
structure remains laminar in this stage. 

Then, the substructure experiences the first turnaround and falls back
again. If the gravitational reaction of the substructure to the larger
cluster is negligible (or, if the test particle assumption is very good), the
substructure keeps oscillating with a constant amplitude
in the larger cluster's potential. However, the test particle assumption
is not good for a longer timescale. Because
of dynamical friction, the oscillation amplitude decreases and the
substructure gradually sinks into the larger cluster's
center. Therefore, the situation like the sloshing model occurs.
As we showed in \S\ref{s:resu}, both Kelvin-Helmholtz and Rayleigh-Taylor
instabilities grow well, and the substructure is destroyed and mixed into
the ambient gas. In this stage, the flow structure becomes highly
turbulent.

\section{Comparison with 1E 0657-56 and A168}
\label{s:comp}

\subsection{1E 0657-56}

1E0657-56 is one of the most well-known examples of a merging cluster.
It is the hottest known cluster and has a very powerful radio
halo \citep[][]{Lian00}.
There are two peaks in both the X-ray
surface brightness distribution \citep[][]{Tuck98,Mark02}
and galaxy distribution \citep[][]{Barr02}, but
their positions do not agree with each other.
Observations of the line-of-sight
velocities of the member galaxies suggest
that its collision axis is almost perpendicular to the line-of-sight
\citep[][]{Barr02}.
\cite{Mark02} shows the detailed analysis of the 
{\it Chandra} observations. A clear cold front and possible bow shock 
are found in the west of the western smaller X-ray peak. Temperature is higher
between the cold front and possible bow shock.

Figure \ref{fig:xrsbewt1} shows the X-ray surface brightness
distribution (contours) and emissivity-weighted temperature map (colors)
of the radial
infall model at $t=0.89$ Gyr. The surface brightness contours are spaced
by a factor of 2.2. We assume that the volume emissivity is 
proportional to the $\rho_{\rm g}^2 T_{\rm g}^{1/2}$, 
where $\rho_{\rm g}$ and $T_{\rm g}$ are the gas
density and temperature, respectively. The line of sight is 
perpendicular to the substructure motion, and the integrations are
carried out within the simulation box. In this simulated X-ray image,
we can see two discontinuities in front of the X-ray peak. The 
emissivity-weighted temperature is higher between those
discontinuities, and is the lowest at the X-ray peak itself.
At the discontinuity nearer to the peak, the emissivity-weighted temperature is
lower in the brighter side, and vice versa. This means that the contact
discontinuity in front of the substructure will be recognized as a cold
front in the X-ray observations.
As for the overall ICM structures of 1E0657-56 around the west X-ray peak, 
our results nicely agree with the X-ray observations.

\cite{Clow04} investigated the mass distribution of 1E0657-56 by
weak gravitational lensing. They show clear offsets of the mass
density peaks from the X-ray ones, and that the mass distribution is well
similar to the galaxy one. The smaller substructure in mass is
ahead of the X-ray one. They argue that this structure occurs because
ICM experiences ram pressure but DM and galaxies do not during cluster mergers.
In contrast with this, our results hardly show such an offset
of the X-ray peak from the potential center that is always at the
center of the simulation box. Evidently, it is necessary to search wider
parameter space in order to clarify if their interpretation is valid.
There might be a feasible parameter set to realize such a situation. 
However, this is not within a scope of this paper.
We should note that the dynamical evolution of the dark matter component is not 
fully calculated self-consistently in our simulations. As a result, the
gravitational potential of the substructure is not modified at all.
Obviously, this is not true in actual clusters. This likely affects
the condition of ram pressure stripping. It will be useful to search
wider parameter space with high resolution N-body +
hydrodynamical simulations.

\subsection{A168}

A168 is a cool irregular cluster. Its X-ray structure is 
highly elongated \citep[][]{Ulme92,Jone99}, 
and two galaxy subclusters are located in its
ends. Recently, \cite{Hall04} revealed the detailed ICM density
and temperature structures around the northern subcluster. 
They found the coolest ICM component 
in the north of the northern subcluster. 
This structure is qualitatively
different than what is found in 1E0657-56. They also show that the tip
of the coolest component consists of a cold front and that its edge is
not so sharp as other examples \citep[e.g.,][]{Mark00,Vikh01b}.

\begin{figure}
\includegraphics[angle=270,scale=1.0]{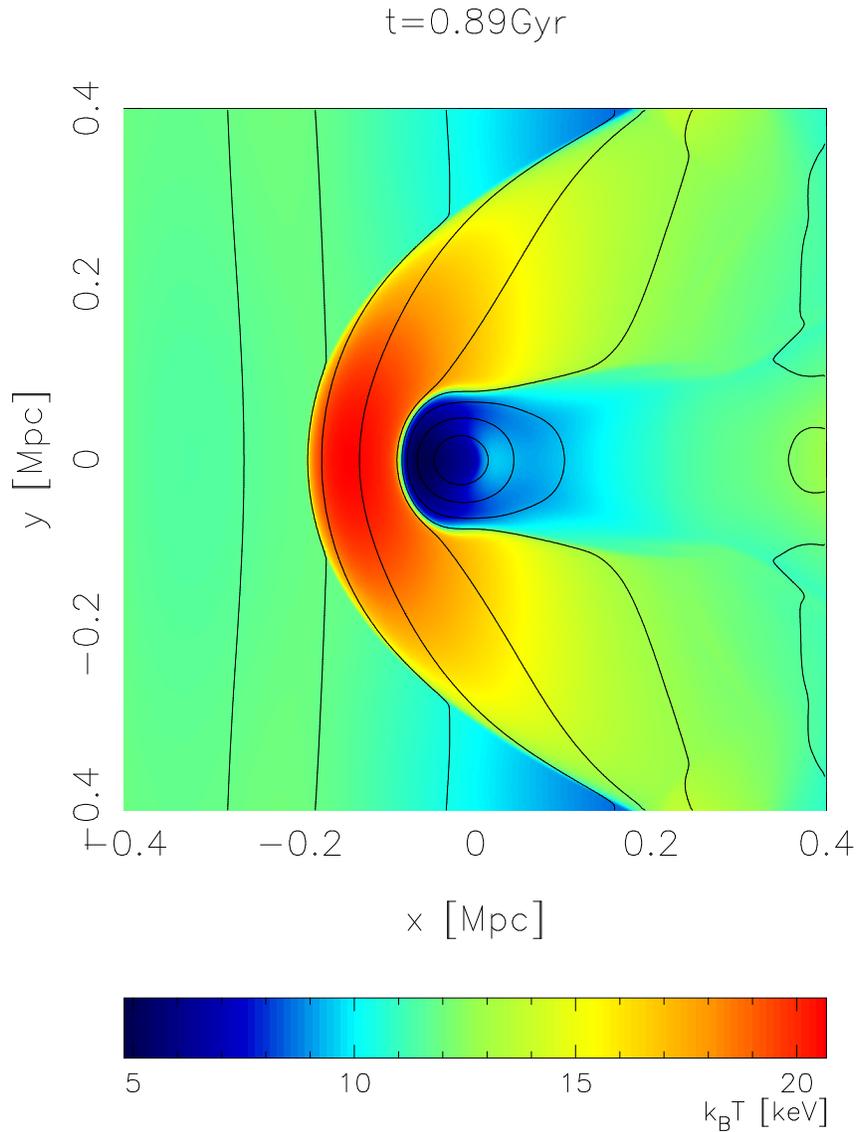}
\caption{The X-ray surface brightness distribution (contours) and
 emissivity-weighted temperature map (colors) for the radial infall model at
 $t=0.89$ Gyr. The surface brightness contours are spaced by a factor
 of 2.2. The line-of-sight is assumed to be perpendicular to the
 substructure motion.}
\label{fig:xrsbewt1}
\end{figure}

\begin{figure}
\includegraphics[angle=270,scale=1.0]{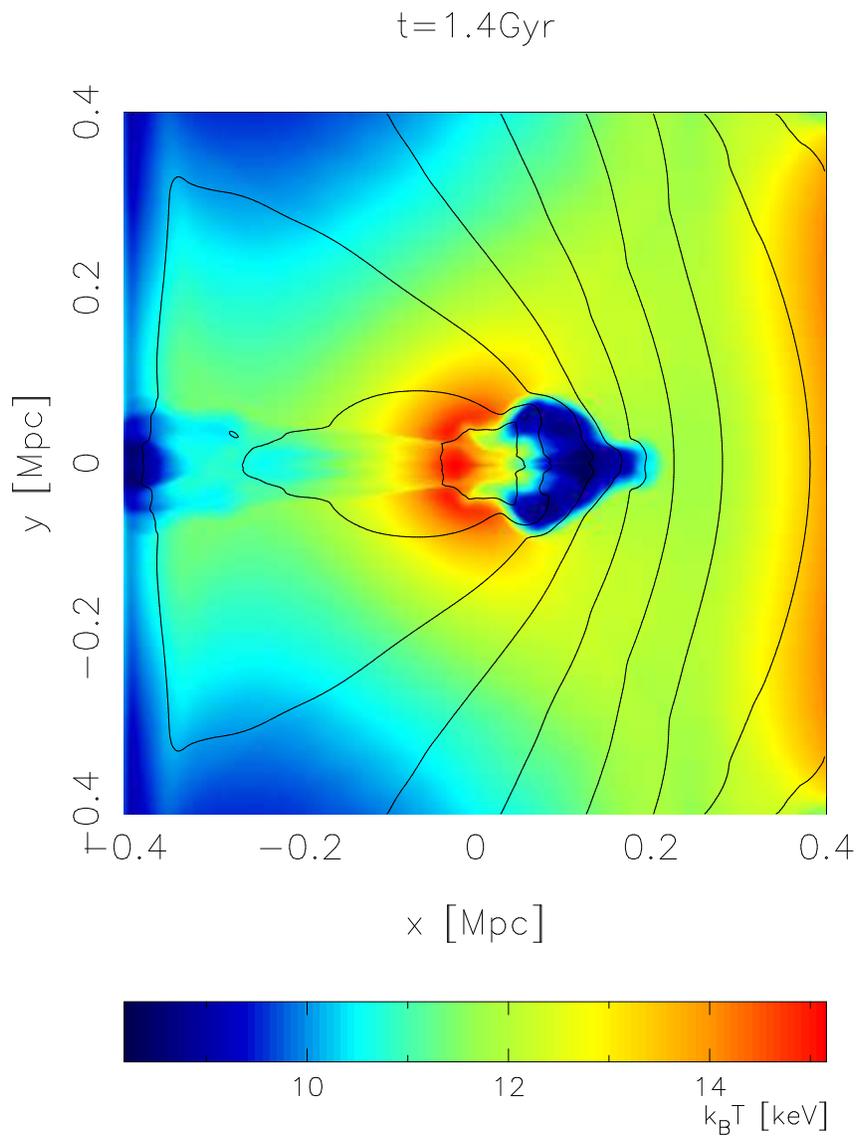}
\caption{The same as in figure \ref{fig:xrsbewt1}, but for the sloshing
 model at $t=1.4$ Gyr. The surface brightness contours are spaced by
 a factor of 1.6.}
\label{fig:xrsbewt2}
\end{figure}

Figure \ref{fig:xrsbewt2} shows the same as in figure \ref{fig:xrsbewt1},
but for the sloshing model at $t=1.4$ Gyr. The surface brightness
contours are spaced by a factor of 1.6. The coldest component of the ICM
appears to be located in front of the X-ray peak. This configuration is
qualitatively the same as what is found around the northern
subcluster of A168. At the front edge of the cold component, the denser
side is colder, and vice versa. Thus, this will be recognized as a cold
front in actual X-ray observations. In addition, this edge is not so sharp as
that in figure \ref{fig:xrsbewt1}, which is consistent with the {\it Chandra}
results of A168 \citep[][]{Hall04}. This is because the contact
discontinuity itself is being perturbed through Rayleigh-Taylor
instabilities. As figure \ref{fig:dentemp2} shows, the cold component in A168 
would be break into pieces and mixed into the ambient ICM in the future.

\section{Summary and Discussion}
\label{s:summ}

We investigate the gas motion in and around a moving substructure in a
cluster of galaxies using the three dimensional Roe-TVD hydrodynamic code.
The subcluster's relative motion in the larger cluster is calculated
assuming that the subcluster is a test particle in the larger cluster's
gravitational potential. We take account of the time variation of the
ICM density and pressure in front of 
the subcluster, and the velocity relative to the larger cluster. Two
representative cases are calculated. One is a simple radial infall of
the subcluster (the radial infall model). The other is a radial oscillation
around the larger cluster's center (the sloshing model). 

In the radial infall model, a clear bow shock and contact discontinuity
form in front of the subcluster. Kelvin-Helmholtz instabilities hardly 
develop because the boundary between the substructure and the
surrounding ICM is not stable because of the ram pressure stripping. As
a result, the flow structure remains laminar before the first turnaround.
In the sloshing model, on the other hand, the flow structure becomes
highly turbulent. In addition to Kelvin-Helmholtz instabilities,
Rayleigh-Taylor instabilities develop well around the turnaround.
 
We compare our results with {\it Chandra} observations of 1E0657-56 and A168.
The X-ray surface brightness distribution and emissivity-weighted
temperature map synthesized from the simulation data of the radial infall
model are in quite good agreement with the X-ray data of 1E0657-56. In
contrast to the weak lensing analysis, however,
the offset of the X-ray peak from the DM potential well is not seen in our
results. As for A168, the X-ray data synthesized from the radial
infall model 
qualitatively agree with the {\it Chandra} results. The relatively blunt cold
front in A168 is probably because of a contact discontinuity perturbed by 
Rayleigh-Taylor instabilities.

In this work, we neglected the magnetic field in the ICM. However, it
is possible that the magnetic field plays a crucial role to suppress the
development of Kelvin-Helmholtz instabilities \citep[][]{Asai04}.  
The magnetic field along the boundary layer probably works to maintain
the smaller subcluster's gas to be a distinct structure
after it is stripped off the DM potential.
Dynamical motion of the substructure
itself possibly produces such kind of magnetic field configurations
\citep[][]{Vikh01a,Asai04}.
It will be useful to perform three dimensional high resolution
magnetohydrodynamical simulations
in order to investigate detailed evolution.
In addition, temperature gradients in the boundary layer possibly produce
the magnetic field structure
through plasma instabilities \citep[][]{Okab03}.



\acknowledgments

Numerical computations were carried out on VPP5000 at the Astronomical
Data Analysis Center, ADAC, of the National Astronomical Observatory of
Japan.
M. T. was supported in part by a Grant-in-Aid from the Ministry
of Education, Science, Sports, and Culture of Japan (16740105)

\clearpage



\end{document}